# Photodetachment of $H^-$ in a quantum well

Guangcan Yang , Kui-Kui Rui and Yizhuang Zheng
School of Physics and Electronic information, Wenzhou University, Wenzhou,325027,China

## Abstract

The photodetachement of $H^-$ in a quantum well is investigated based on closed orbit theory. It is found the distances between the ion and the two hard walls modulate the cross section of photodetachment. For the hard walls perpendicular the polarization of photons, the detachment spectrum displays a staircase structure. The pattern of staircase is modulated by the ratio of distances from the ion to the two walls while the distance controls step interval of the staircase structure.

The interaction between atoms[1], ions[2] and molecules[3] with metal and dielectric surfaces has attracted considerable interest in recent years. The interaction between the atomic and metallic electronic clouds can lead to a variety of phenomena, such as charge transfer and associated orbital hybridization processes. Aside from their basic scientific relevance, studies of atom(molecule)-surface interaction become increasingly important in conjunction with technological applications such as the development of ion sources, control of ion-wall interactions in fusion plasma, surface chemistry and analysis, secondary-ion mass spectroscopy, reactive ion etching, and semiconductor miniaturization via thin-film deposition[4,5].

On the other hand, the photo-induced electronic excitation of atoms and ions on metal surfaces interact with the surface states and its dynamical evolution reveals important imformation on the surface and its adsorbates[6,7]. Rencently, we investigated the photodetachment of $H^-$ with and without an electric field near an elastic wall based on the semiclassical closed orbit theory and "exact" quantum approach[8,9,10]. It has been found that the cross section of near an elastic wall shows an irregular staircase structure with sharp edge of each step. The same system without the external electric field shows a sinusoidal modulation in the cross section. In present work, we put a negative ion of hydrgen in a quantum well consisting of two hard walls and investigate its photo-induced properties by closed orbit theory[11,12]. We find that its photodetachment cross section show a similar staircase curve but with a complicated structure which is related the ratio of distances from the ion to the two walls. Such interesting structure can be attributed to the interference between two beams of photodetached electronic waves in opposite directions.

We consider the following system: In the cylindrical coordinates ($\rho, z, \phi$), a hydrogen negative ion $H^-$ sits at the origin and a $z$-polarized laser is applied for the photodetachment. The ion is in a quantum wall where two hard wall locate at $z_1$ (up wall) and $-z_2$ (down wall) respectively(we take positive $z_1$ and $z_2$). Thus the photodetached electrons move freely in $(\rho, \phi)$ plane and are bounced back and forth by the two walls in $z$-direction. In all the classical trajectories of the photodetached electron emanating out the origin, those bounced back by the walls to the starting point are called closed orbits. The system of closed orbits is similar to the case of the gradient electric field[13]. Clearly, all the closed orbits lay in $z$-axis, and the following are fundamental: ( All other closed orbits consist of these fundamental orbits and their repetitions). (i) The electron goes up along the +$z$ direction, reaches its maximum and then bounced back by the up wall and return to the origin. We call this orbit the up orbit. (ii) The electron goes down in –$z$ direction and hits the down wall and bounces back and finally return to the origin. We call this orbit the down orbit. (iii) The electron completes the up orbit first and then passing through the origin, and continues to complete the down orbit. (iv) The electron completes the down orbit first and then up orbit. This orbit is similar to the one of (iii) but in reverse order.

To classify the system of closed orbits, indices $j$ and $n$ are used to label the orbits, where $j=1,2,3,4$ and $n=0,1,2,3…$. Here $n=0$ means the orbit is fundamental closed orbit( $j=1,2,3$ and 4 for the fundamental closed orbits above, respectively). When n>0, orbit($j, n$) has two parts, the beginning part and the later part. The beginning part is always the $j$th fundamental closed orbit. The later part consists of n repetitions of periodic orbits $j=3$ or 4. For the case of $j=1$, it's $n$ repetitions of $j=3$, and for $j=2$, it corresponds to $n$ repetitions of $j=4$. For periodic orbit $j=3$ or $j=4$, it repeats itself $n$ times. The returning time $T_j$ of the fundamental closed orbits are

$$\begin{aligned} T_1 &= \frac{2z_1}{k} \\ T_2 &= \frac{2z_2}{k} \\ T_3 &= \frac{2(z_1 + z_2)}{k} \\ T_4 &= \frac{2(z_1 + z_2)}{k} \end{aligned} \quad (1)$$

Clearly, we have $T_3 = T_4 \equiv T$. The classical actions along the fundamental closed orbits are

$$\begin{aligned} S_1 &= 2\sqrt{2}E^{1/2}z_1, \\ S_2 &= 2\sqrt{2}E^{1/2}z_2, \\ S_3 &= S_1 + S_2, \\ S_4 &= S_1 + S_2. \end{aligned} \quad (2)$$

Clearly, we have $S_3 = S_4 \equiv S$. Maslov indices of these orbits can be easily found by counting the returning points where a $\pi$ phase loss of electronic wave occurs, thus we have

$$\begin{aligned} \mu_1 &= \mu_2 = 2, \\ \mu_3 &= \mu_4 = 4. \end{aligned} \quad (3)$$

The action of any closed orbit consists of the ones of fundamental orbits and can be written as

$$S_{jn} = S_j + nS. \quad (4)$$

The returning time of any closed orbit is

$$T_{jn} = T_j + nT \quad (5)$$

with Maslov index

$$\mu_{jn} = \mu_j + 4n. \quad (6)$$

For hydrogen ion H⁻ in a quntum well, the photodetachement can be regarded as a one-electron process if we neglect the influence of the short-range potential $V_b(r)$ when the electron is far from the core. The binding energy $E_b = k_b^2/2$ is approximately 0.754 eV, where $k_b$ is related with the initial wave function $\Psi_i = C\exp(-k_b r)/r$, $C$ is a "normalization function" constant and is equal to 0.31552[6]. When an incident laser beam is applied to the ion the valence electron absorbs a photon energy $E_p = E_b + E$ and the steady outgoing electron wave propagates outward in all directions. The electron moves in a straight line at constant speed before it hits the hard walls. We propagate the electron wave semiclassically. The semiclassical wave is finally reflected by the elastic walls. Thus, some of the wave returns to the vicinity of the core and interfere with the outgoing wave to produce the spectral pattern of the photodetachment[11,12]. In this physical picture, the photodetachment cross section can be expressed in the form[14]

$$\sigma(E) = -\frac{4E_p}{c}\text{Im}\left\langle D\Psi_i \left| \hat{G}^+ \right| D\Psi_i \right\rangle \qquad (7)$$

The outgoing Green's function is denoted by $\hat{G}^+$. The dipole operator $D$ is equal to the projection of the electron coordinate onto the direction of polarization of the laser field. Eq.(7) has the following physical meaning: the initial state is modified by the dipole operator related with the incident laser field to become the source wave function; the Green's function propagates these waves outward to become the outgoing waves; and finally the waves overlap with the source wave to give the absorption spectrum.

The outgoing wave can be divided into two parts. The first part never goes far from the core and it is called the direct wave. The second part is known as the returning wave that propagates outward into the external region first, then is reflected by the elastic walls, and finally returns to the vicinity of the core to interfere with the outgoing wave. Accordingly, the cross section has two parts,

$$\sigma(E) = \sigma_0(E) + \sigma_{ret}(E) \qquad (8)$$

where the field-free cross section $\sigma_0(E) = \frac{16\sqrt{2}C^2\pi^2}{3c}\frac{E^{3/2}}{(E_b+E)^3}$ is the contribution of the direct wave interfering with the source, and the second part $\sigma_{ret}(E) = -\frac{4E_p}{c}\text{Im}\left\langle D\Psi_i \middle| \Psi_{ret} \right\rangle$ is the contribution of the returning wave overlapping with the outgoing source wave. We calculate the returning wave by semiclassical approach.

To construct the semiclassical wave function we start from a spherical surface centered at the origin with radius $R \approx 10a_0$. The initial outgoing wave on the surface of this sphere is

$$\Psi^0(R,\theta,\phi) = -i\frac{4Ck^2}{\left(k_b^2+k^2\right)^2}\cos(\theta)\frac{e^{i(kR-\pi)}}{kR}, \tag{9}$$

then the wave propagating from the sphere can be expressed semiclassically as

$$\Psi(\rho,z,\phi) = \sum_i \Psi^0(R,\theta,\phi) A_i e^{i[S_i-\mu_i\pi/2]} \tag{10}$$

where $S_i$ is the action along the $i$th trajectory, $\mu_i$ is the Maslov index characterizing the geometrical properties of the $i$th trajectory and its neighboring orbits. The amplitude of the wave function, $A_i$ is given by

$$A_i(\rho,z,\phi) = \left|\frac{J_i(\rho,z,0)}{J_i(\rho,z,t)}\right|^{1/2} = \frac{R}{R+kt} \tag{11}$$

where

$$J(\rho,z,t) = \rho(t)\begin{vmatrix} \frac{\partial z}{\partial t} & \frac{\partial z}{\partial \theta} \\ \frac{\partial \rho}{\partial t} & \frac{\partial \rho}{\partial \theta} \end{vmatrix}. \tag{12}$$

Only the waves associated with the closed orbits return to the vicinity of the core and interfere with the steadily-producing outgoing spherical waves and therefore contribute the cross section. The returning waves near the core must be cylindrically symmetric and thus can be approximated by incoming Bessel functions. And the returning wave function related with closed orbit ($j,n$) must match the Bessel function and can be written as

$$\Psi_{jn}^{ret} = N_{jn}\frac{1}{\sqrt{2\pi}}J_0(k_\rho^{ret}\rho)\frac{1}{\sqrt{2\pi}}e^{ik_z^{ret}z}, \tag{13}$$

where the normalization factor $N_i$ can be determined by matching Eq.(10) and Eq.(13). For closed orbit ($j,n$), the modified amplitude in Eq. (11) of the returning function is

$$A_{jn} = \frac{1}{T_{jn}k} \tag{14}$$

then the matching factor in Eq. (13) becomes

$$N_{jn} = (-1)^{\mu_j/2-1}\frac{4iCk^2}{T_{jn}k\left(k_b^2+k^2\right)^2}\exp\left[\frac{i}{y}\left(S_j-\mu_j\frac{\pi}{2}\right)\right] \tag{15}$$

The returning part of the cross section can now be calculated by overlapping the returning waves with the outgoing wave functions, then taking the imaginary part. Therefore, we have

$$\sigma_{ret}(E) = -\frac{4E_p}{c}\sum_{jn} \text{Im}\int C\exp(-k_b r)\cos\theta\, N_{jn} e^{-ikr\cos\theta_{jn}^{ret}} r^2 dr \sin\theta\, d\theta\, d\phi, \quad (16)$$

Integrating out the expression and adding the field free part, we have the total photodetachment cross section

$$\sigma(E) = \frac{16\sqrt{2}C^2\pi^2}{3c}\frac{E^{3/2}}{(E_b+E)^3} + \sum_{jn} C_{jn}\sin(S_{jn} - \mu_{jn}\pi/2). \quad (17)$$

where $c$ is light speed and

$$C_{jn} = (-1)^{\mu_j/2-1}\frac{2\pi^2}{c}\frac{2\sqrt{2}C^2 E^{1/2}}{T_{jn}(E_b+E)^3}. \quad (18)$$

The photodetachment cross section can be calculated for different values of the distances between the ion and the two elastic walls and different ratio of them. The results are shown from Fig. 1 to 3. We can see that the detachment spectrum displays a staircase structure. This structure results from the superposition of the contribution from the four fundamental closed orbits and their repetitions. In Fig.1, we keep the distance to the down wall $z_2$ fixed at 200 a. u. and change $z_1$, the distance to the up wall. When the distance $z_1$ is very large, the present system should recover the case when only one hard wall exists. Taking $z_1 = 6000$ a. u., the calculated cross section is shown in Fig. 1(a). The spectrum is almost the same as a superposition of a background and a sinusoidal oscillation except some very small steps appeared in the oscillation. When $z_0 \to \infty$, then $T_2$, $T_3$, and $T_4$ tend to infinity, so there are no contributions from these closed orbits, the only closed orbit makes the sinusoidal spectral oscillation. When the closer we move the wall to the ion, the more contribution from the orbits bounces back by the up wall. We can see the elastic wall change the cross section from the smooth oscillation to a staircase curve. If the wall approaches the ion further, we can see the interval of step of the staircase goes wider in Fig. 1(b) and (c).

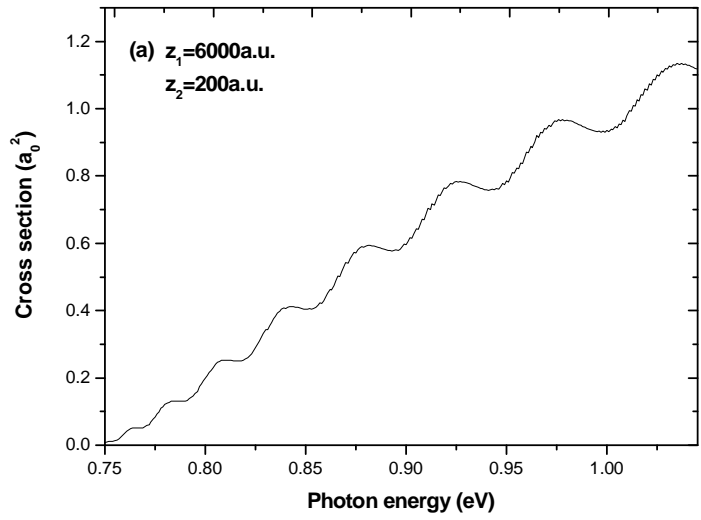

Fig. 1(a)

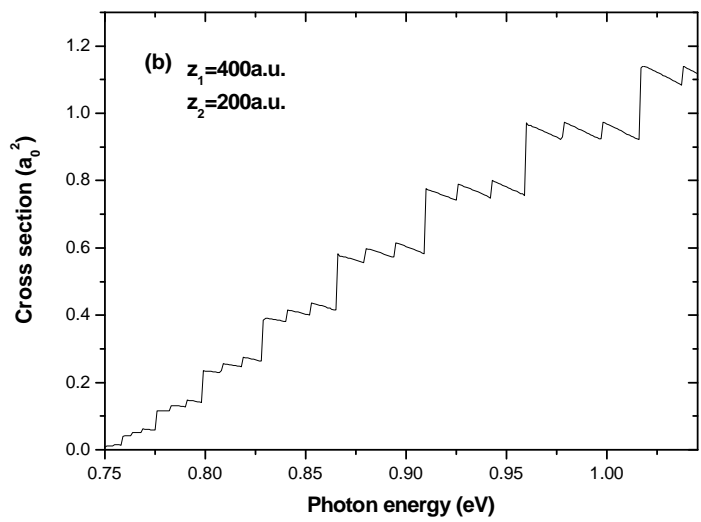

Fig. (b)

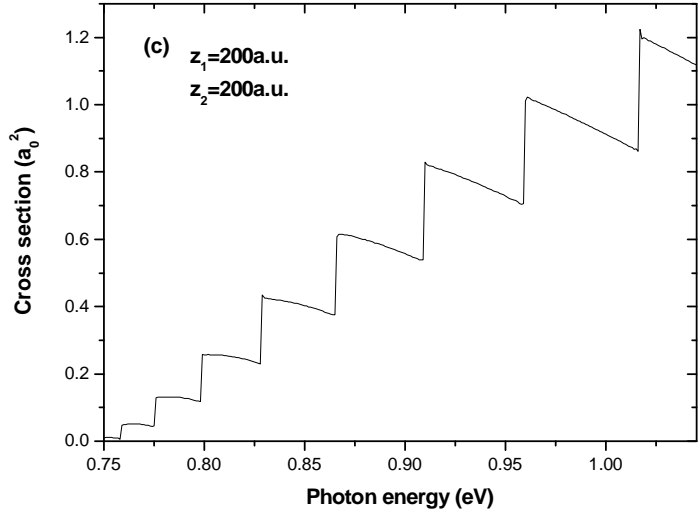

Fig. 1(c)

In Fig.2, we show how the ratio of the two distance to walls influence the pattern of the staircase. When the ratio is a rational number, we have a periodic structure as shown in Fig. 2(a) where $z_1/z_2 = 1.5$. If the ratio is irrational, the periodic structure is replaced by an irregular staircase curve, as shown in Fig. 2(b) and (c) with $z_1/z_2 = \sqrt{2}, \sqrt{3}$ respectively. This phenomenon can be understood by the interference of closed orbits in Eq.(17). When the phases of closed orbits are matched, they interfere constructively and make a regular pattern. While they are mismatched, a irregular staircase pattern appears.

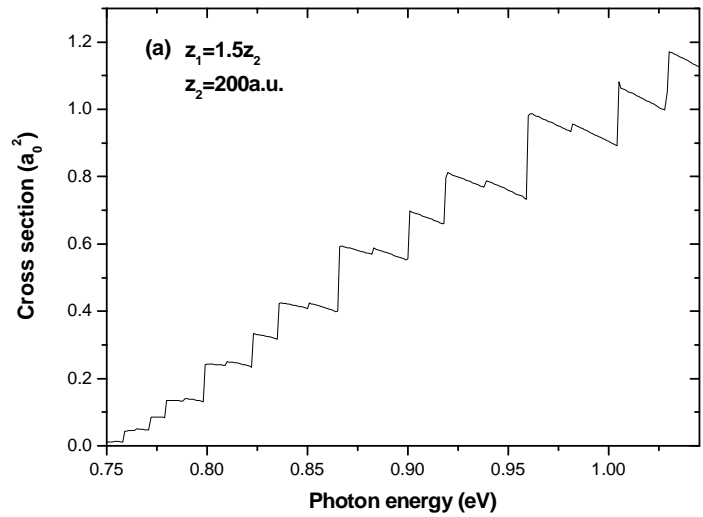

Fig.2(a)

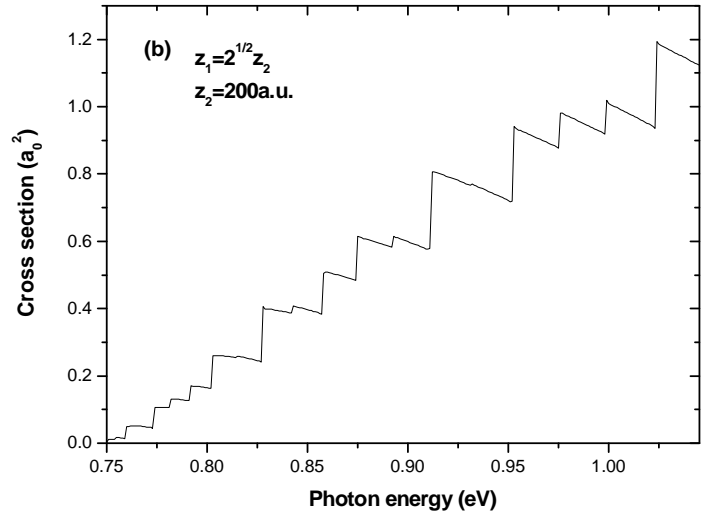

Fig.2(b)

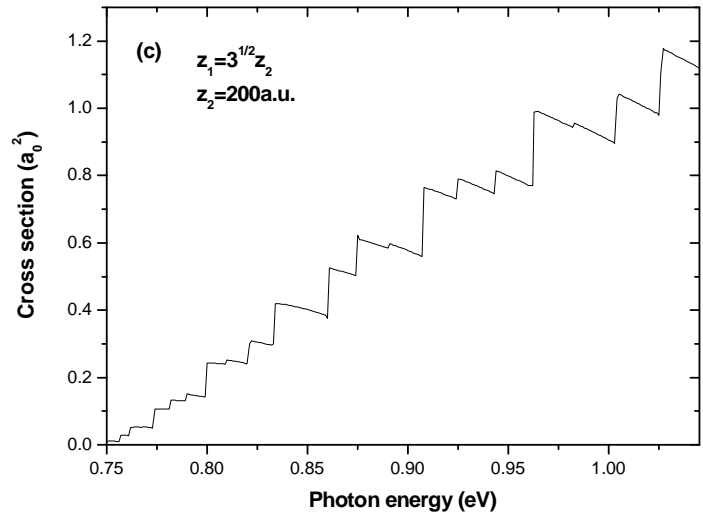

Fig.2(c)

We explore this interesting feature further by fixing the ration but change the two distances to the walls simultaneously. The result is shown in Fig. 3. In Fig. 3(a) and (b), the ratio is fixed to 2 but $z_2$ goes from 200 a.u. to 400 a.u. We can see that the two pattern are similar but with different interval of step of staircase. Fig. 3(c) and (d) show the case of the irrational ratio. We can see the obvious similarity between the two irregular patterns but no perodity.

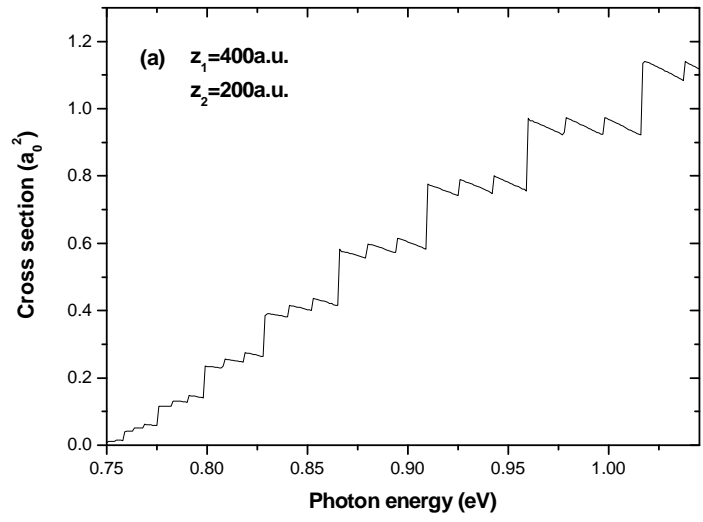

Fig. 3(a)

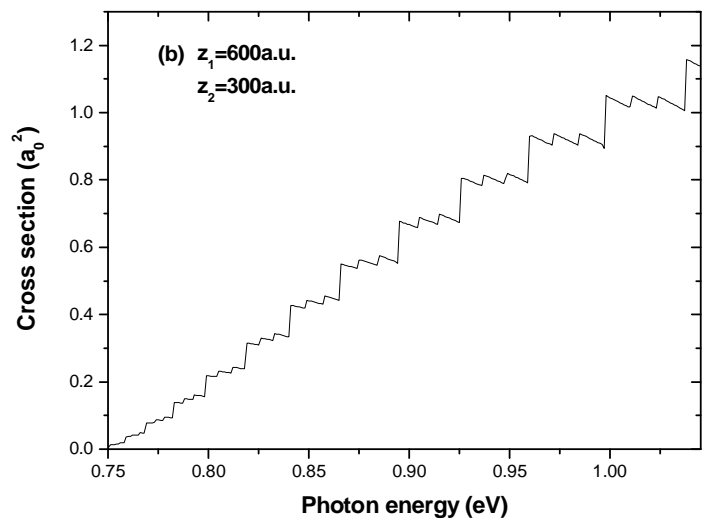

Fig. 3(b)

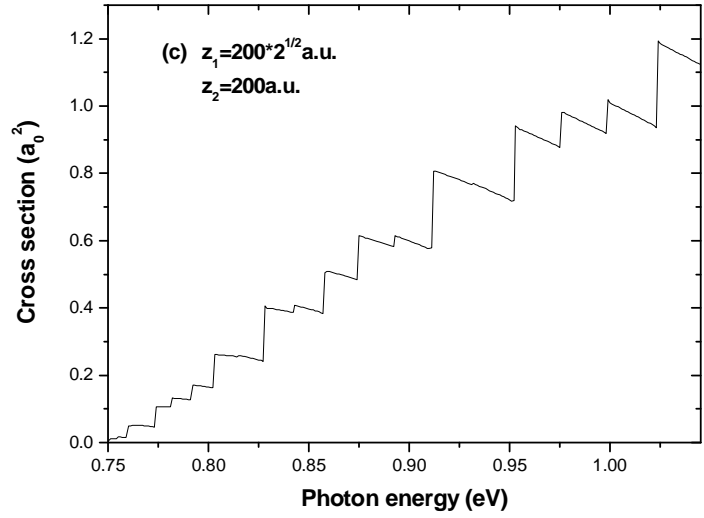

Fig. 3(c)

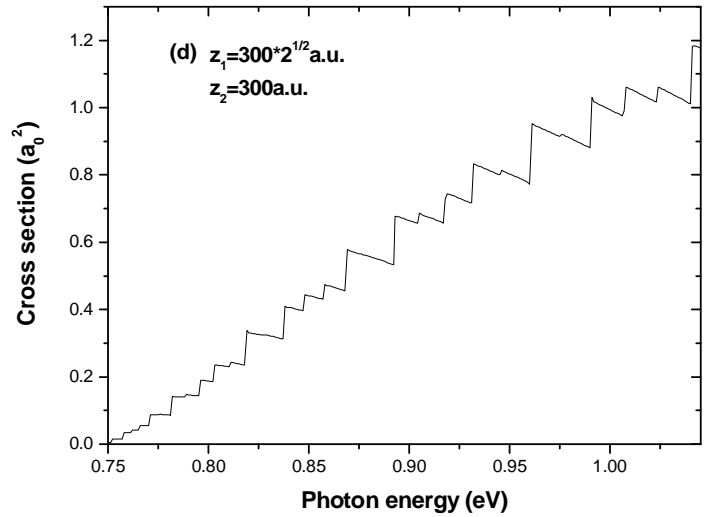

Fig.3(d)

In summary, the photodetachment cross section of H⁻ in a quantum well has been investigated in the framework of closed orbit theory and an analytical cross section has been obtained as the summation of all closed orbits. We found that the photodetachment spectrum shows staircase structure because of the constraint of the quantum well to the photodetached electrons, contrasting with the regular smooth oscillating curve when only an elastic wall exists. The new feature in the present system is that the ratio modulates the pattern of the staircase while the distances determine the step interval of staircase.

# FIGURE CAPTIONS

Figure 1: The dependence of the photodetachment cross section on the distance between the negative hydrogen ion and the up wall. The distance to the down wall is fixed at 200 a. u. and the distance is (a) $z_1 = 6000$ a.u., (b) $z_1 = 400$ a.u., and (c) $z_1 = 200$ a.u.

Figure 2: The dependence of pattern of the photodetachment cross section on ratio of distances to wall. The distance between the ion and the down wall is fixed at $z_2 = 200$ and the ratio $z_1/z_2$ is (a)1.5, (b) $\sqrt{2}$, and (c) $\sqrt{3}$.

Figure 3. Comparison between different distances with a fixed ratio. (a) $z_1/z_2 = 2$, $z_2 = 200$ a.u. (b) $z_1/z_2 = 2$, $z_2 = 300$ a.u. (c) $z_1/z_2 = \sqrt{2}$, $z_2 = 200$ a.u. (d) $z_1/z_2 = \sqrt{2}$, $z_2 = 300$ a.u.